\documentclass[preprint,12pt]{elsarticle}

\usepackage{amssymb}
\usepackage{multirow}
\usepackage{booktabs}
\usepackage{graphicx}
\usepackage{float}
\usepackage{amsmath}

\journal{Applied Soft Computing}

\begin{document}

\begin{frontmatter}



\title{Uncertainty-aware Generative Learning Path Recommendation with Cognition-Adaptive Diffusion}


\author[inst1]{Xiangrui Xiong} 
\author[inst1]{Hang Liang}
\author[inst1]{Baiyang Chen}
\author[inst1]{Zifei Pan}
\author[inst1]{Yanli Lee\corref{cor1}}
\cortext[cor1]{Corresponding author}
\ead{yanlicomplex@gmail.com}

\affiliation[inst1]{organization={Xihua University},
            addressline={School of Computer and Software Engineering},
            city={Chengdu},
            postcode={610039},
            country={China}}

\begin{abstract}
Learning Path Recommendation (LPR) is critical for personalized education, yet current methods often fail to account for historical interaction uncertainty (e.g., lucky guesses or accidental slips) and lack adaptability to diverse learning goals. We propose U-GLAD (\textbf{U}ncertainty-aware \textbf{G}enerative \textbf{L}earning Path Recommendation with Cognition-\textbf{A}daptive \textbf{D}iffusion). To address representation bias, the framework models cognitive states as probability distributions, capturing the learner’s underlying true state via a Gaussian LSTM. To ensure highly personalized recommendation, a goal-oriented concept encoder utilizes multi-head attention and objective-specific transformations to dynamically align concept semantics with individual learning goals, generating uniquely tailored embeddings. Unlike traditional discriminative ranking approaches, our model employs a generative diffusion model to predict the latent representation of the next optimal concept. Extensive evaluations on three public datasets demonstrate that U-GLAD significantly outperforms representative baselines. Further analyses confirm its superior capability in perceiving interaction uncertainty and providing stable, goal-driven recommendation paths.
\end{abstract}

\begin{graphicalabstract}
    \centering 
  \includegraphics[width=0.95\textwidth]{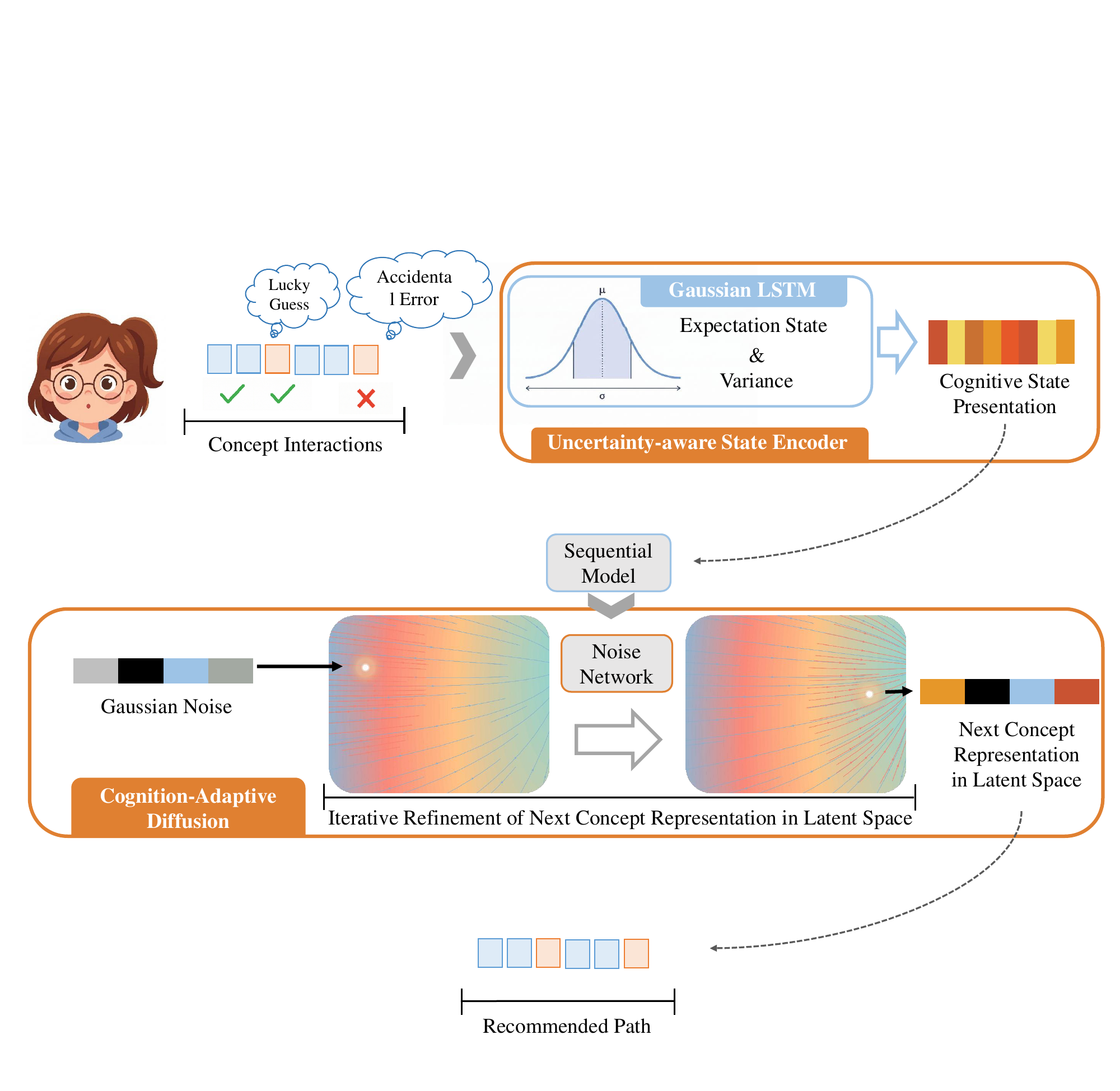}
\end{graphicalabstract}

\begin{highlights}
\item Unify interaction uncertainty and goal-aware encoding in a cohesive LPR framework.
\item Propose U-GLAD with Uncertainty-aware State Encoding and Diffusion Decoding.
\item Gaussian LSTM handles interaction noise for robust cognitive state modeling.
\item U-GLAD achieves state-of-the-art results across three educational datasets.
\end{highlights}


\end{frontmatter}


\section{Introduction}
\label{intro}
With the rapid proliferation of online education platforms and the explosive growth of digital learning resources \cite{SunChen2016, CastroTumibay2021, DumfordMiller2018}, providing personalized learning experiences has become a core demand in educational informatization \cite{Bernackietal2021, Ayenietal2024}. Against this backdrop, Learning Path Recommendation (LPR), as a pivotal educational auxiliary technology, has garnered significant attention in recent years\cite{NabizadehEtal2020, RahayuEtal2023, shi2020learning}. The primary objective of LPR is to formulate effective concept sequences tailored to a learner’s cognitive state and specific learning goals. By organizing discrete concepts into coherent learning paths, advanced LPR systems can significantly enhance cognitive levels while mitigating learning fatigue \cite{HuangEtal2019, ChenEtal2023, LiEtal2023, ZhangEtal2024, LuoEtal2025, YuEtal2025, ChengEtal2026}.

Regarding cognitive state modeling, existing methods \cite{ChenEtal2023, YuEtal2025, ZhangEtal2024} typically employ sequential models to fit historical interaction sequences, yielding deterministic representations of cognitive states. However, a critical issue (RQ1) remains: historical interaction sequences are not entirely reliable, as they are often confounded by uncertainties arising from random factors such as lucky guesses or accidental slips \cite{ChengEtal2025} . Consequently, over-fitting to these sequences leads to biased estimations of the true knowledge state, which subsequently impairs the decoding process. In terms of concept representation, DLPR \cite{ZhangEtal2024} pioneered the integration of item difficulty into concept embeddings to ensure that recommended concepts match the learner’s current level, thereby addressing the issue of "rugged" paths caused by drastic fluctuations in difficulty. Subsequently, SRC \cite{ChenEtal2023} utilized the Set-to-Sequence paradigm \cite{VinyalsEtal2015} to capture prerequisite relations among concepts, obtaining more logically rigorous representations that improve the performance of pointer networks during next-concept decoding. Building upon this, LIGHT \cite{YuEtal2025} further explored prerequisite and synergistic relationships from a graph-based perspective to enrich concept semantics. Nevertheless, the concept representations produced by these methods lack essential learning goal information (RQ2).

Furthermore, during the decoding stage, conventional schemes utilize pointer networks as discriminative models to select the next concept from a candidate set. Despite the competitive results of discriminative decoding, research \cite{LiEtal2023DiffuRec} suggests that generative solutions based on diffusion models \cite{ho2020denoising} can effectively mitigate historical interaction uncertainty to build superior recommendation systems. We argue that LPR is particularly well-suited for generative approaches because the recommended path should not merely fit historical data; rather, it should be an idealized sequence generated from the cognitive state distribution to maximize the mastery of learning goals. Therefore, providing a generative diffusion model solution to predict the latent representation of the next concept within the LPR domain serves as a primary research objective of this paper (RQ3).

\section{Related Work}
\label{rw}
\subsection{Learning Path Recommendation}
\label{rw-lpr}
Learning Path Recommendation (LPR) aims to curate personalized concept sequences tailored to a learner’s cognitive state and specific learning objectives. Early methodologies relied on heuristic algorithms \cite{birjali2018novel, zare2016multi, DwivediEtal2018, LiuEtal2019, gavrilovic2022design}, such as genetic algorithms \cite{DwivediEtal2018} , or utilized reinforcement learning for rule-based matching \cite{LiuEtal2019} . With the advancement of deep learning, researchers \cite{ZhouEtal2018} began employing sequential models like RNNs \cite{Graves2012} for path generation. Following the rise of self-attention mechanisms, Transformers \cite{vaswani2017attention} have become the predominant representation models in the LPR field due to their superior capability in learning feature representations. Despite varying architectures, recent works \cite{ChenEtal2023, YuEtal2025, ZhangEtal2024, ChengEtal2026} can generally be categorized into two classes:

(1) LPR based on item difficulty and cognitive load constraints. These algorithms focus on providing paths with low cognitive load to minimize learner frustration. DLPR \cite{ZhangEtal2024} conceptualizes LPR from a "real walking" perspective, arguing that ignoring item difficulty leads to inefficient paths. To address this, DLPR integrates difficulty information into concept representations and utilizes a hierarchical reinforcement learning framework to strictly control the difficulty of recommended items. KnowLP \cite{ChengEtal2026} suggests that over-reliance on prerequisite relationships can cause learners to stagnate; thus, it constructs a prerequisite graph to dynamically switch to similar but easier concepts when a learner encounters obstacles.

(2) LPR based on concept structural relationship decoding. These methods typically adopt an encoder-decoder architecture, where the encoder extensively explores structural relations among concepts to provide high-quality representations for the decoding process. SRC \cite{ChenEtal2023} leverages the Set-to-Sequence \cite{VinyalsEtal2015} paradigm and self-attention \cite{vaswani2017attention} to capture prerequisite dependencies within concept sets, significantly improving path generation. Building on SRC \cite{ChenEtal2023}, LIGHT \cite{YuEtal2025} delves deeper into structural dependencies by constructing prerequisite and synergistic graphs, ensuring that concept representations incorporate both sequential dependencies and mutual support information.

While these studies have achieved impressive results by focusing on enriched concept embeddings and discriminative decoding, they have yet to account for uncertainty in historical interactions or explore generative decoding strategies.

\subsubsection{Historical Sequence Uncertainty}
Research \cite{LiEtal2023DiffuRec} in recommendation systems indicates that user history often contains uncertainties. While systems predict future interactions based on historical probability distributions, the presence of erroneous records can cause uncertainty and lead to modeling bias if over-fitted. In the field of Knowledge Tracing (KT) \cite{abdelrahman2023knowledge}, which is closely related to LPR, UKT \cite{LiEtal2023} has confirmed the existence of uncertainty in learner interactions. Factors such as accidental slips or lucky guesses result in false negatives and false positives, which can mislead models in assessing a learner's cognitive state.

To address the neglect of historical sequence uncertainty, UKT adopts stochastic distribution embeddings to decouple fundamental knowledge levels (quantified by the mean) from uncertainty (quantified by covariance). Furthermore, UKT employs contrastive learning with negative samples representing guesses and slips to enhance model robustness. In the broader recommendation domain, DiffRec posits that recommendation is essentially an inference of future interaction probabilities based on "corrupted" historical data. The mechanism of diffusion models—where a forward process injects Gaussian noise to "corrupt" the history and a reverse process recovers the truth—is highly compatible with this recommendation process, allowing for the mitigation of historical uncertainty.

In summary, current LPR methods lack a mechanism to account for historical interaction uncertainty, which negatively impacts the quality of recommended paths. Within the U-GLAD framework, we design a Gaussian LSTM to mitigate such uncertainty, yielding smoother and more authentic cognitive state representations. Furthermore, to further eliminate uncertainty during the decoding stage, we introduce a cognition-adaptive diffusion model to predict the latent representation of the next concept in a generative manner.

\section{Problem Definition}
\label{sec:problem_definition}

In this study, we formalize the knowledge concept structure-based, goal-oriented learning path recommendation task \cite{ChenEtal2023, ChengEtal2026,LuoEtal2025, YuEtal2025, LiEtal2023, ZhangEtal2024}. Let $C = \{c_{1}, c_{2}, \dots, c_{N}\}$ denote the universal set of $N$ knowledge concepts within a specific domain. A learner's specific learning objectives are defined as a subset $G \subseteq C$, where $G = \{g_{1}, g_{2}, \dots\}$ represents the target concepts to be mastered. Each historical interaction is represented as a tuple $h = (c, y)$, consisting of a concept $c \in C$ and a corresponding label $y \in [0, 1]$ that indicates the mastery status. Given a historical sequence $H = \{h_{1}, h_{2}, \dots, h_{i}\}$ and the target goals $G$, our framework aims to recommend an optimal learning path $P = \{\tilde{c}_{1}, \tilde{c}_{2}, \dots, \tilde{c}_{k}\}$, where each $\tilde{c}_j \in C$ denotes a recommended learning object.

To evaluate the effectiveness of the recommended sequence, we utilize the mastery improvement rate ($E_T$) as the primary metric \cite{ChenEtal2023, YuEtal2025}. For a given session, $E_T$ quantifies the educational gain by measuring the relative progress made toward total mastery of the target goals. The improvement rate is formulated as:
\begin{equation}
    E_{T} = \frac{E_{e} - E_{b}}{1 - E_{b}}
    \label{eq:improvement_rate}
\end{equation}
where $E_{b}$ represents the initial mastery level before the learning session, and $E_{e}$ denotes the final mastery level achieved after the learner completes the recommended path $P$. In this context, the value $1$ represents the theoretical maximum mastery level. The ultimate objective of the proposed U-GLAD framework is to generate a sequence $P$ that maximizes the expected improvement rate $E_{T}$.

\section{Method}
SRC \cite{ChenEtal2023} posits that learning path recommendation is essentially a process of arranging a set of concepts into a sequence, which can be addressed through an encoder-decoder architecture to decouple concept representation learning from the sequence generation process. While this architecture achieves high-quality representations of cognitive states and concepts, and utilizes discriminative decoders to improve goal mastery, it often overlooks historical interaction uncertainty. Consequently, we adopt a classic encoder-decoder framework but introduce an Uncertainty-aware State Encoder to handle interaction noise and a Cognition-Adaptive Diffusion model as the primary component of the decoder to predict the latent representation of the next concept in a generative manner. To further enhance personalization, we design a goal-oriented concept encoder that adjusts concept representations based on specific learning objectives. Figure 1 provides an overview of the U-GLAD framework.

\begin{figure}[H] 
\centering
\includegraphics[width=0.90\textwidth]{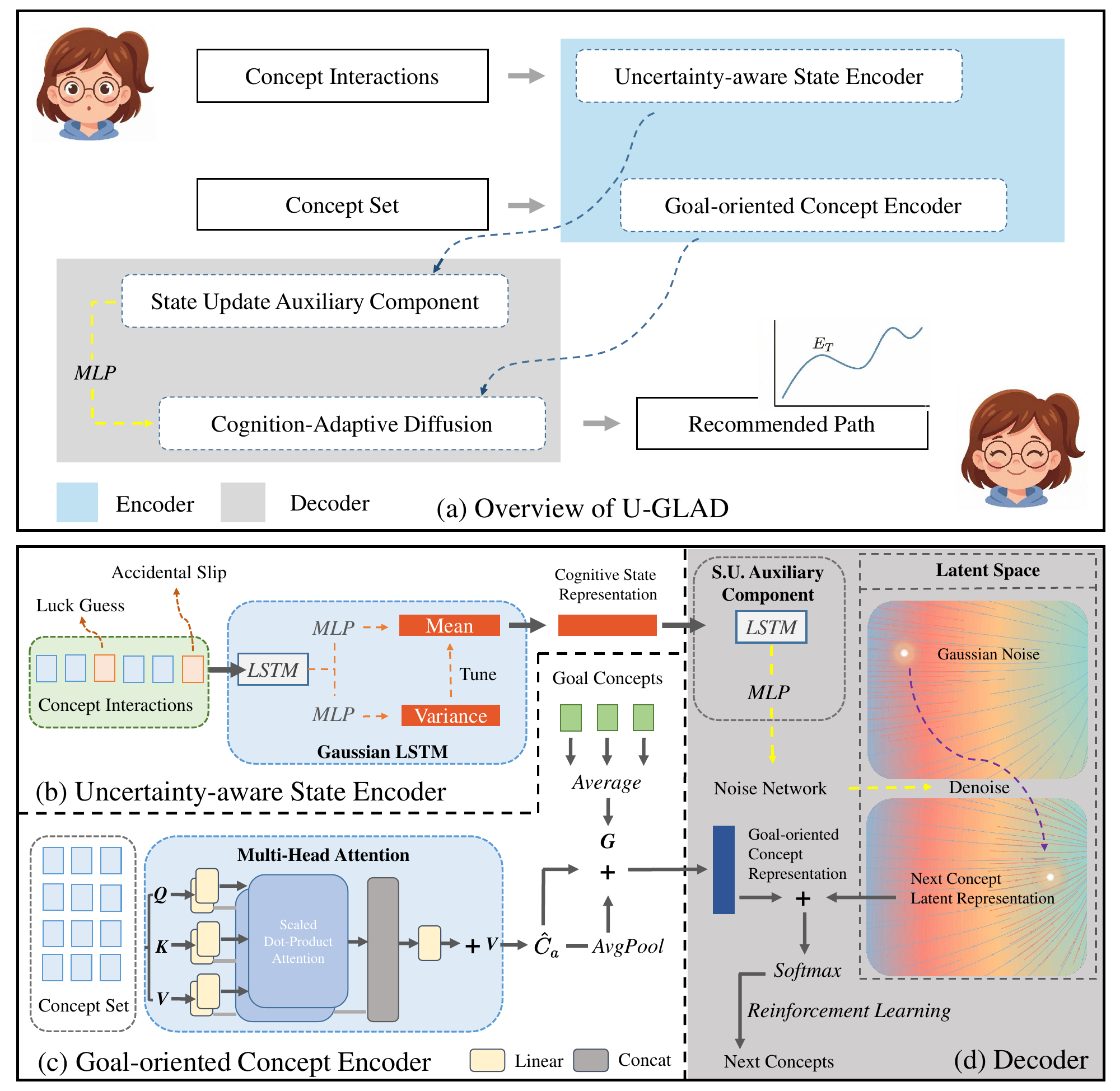} 
\caption{System architecture of U-GLAD. The framework unifies uncertainty-aware state encoding and cognition-adaptive diffusion decoding.}
\label{fig:framework}
\end{figure}

\subsection{Encoder}
The encoder in the U-GLAD framework is designed to produce high-quality representations of both the learner's cognitive state and the knowledge concepts. To mitigate the impact of historical interaction uncertainty and obtain a more authentic cognitive state, we design an uncertainty-aware state encoder using a Gaussian LSTM as the backbone. Furthermore, recognizing that different learners have distinct objectives, we develop a goal-oriented concept encoder to incorporate target information into concept embeddings, facilitating more effective and personalized decoding.

\subsubsection{Uncertainty-aware State Encoder}
Historical interaction uncertainty arises from behaviors such as accidental slips or lucky guesses, making such records unreliable and detrimental to recommendation performance \cite{ChengEtal2025, LiEtal2023DiffuRec}. To address this, our Gaussian LSTM models the cognitive state as a probability distribution, utilizing mean and variance operations to derive a de-noised state representation.

Specifically, the Gaussian LSTM predicts the expected cognitive state, while the variance quantifies the associated uncertainty. The mechanism is described by equations (2-5):
\begin{equation}
\mathbf{h}_{u}^{t}=f_{e}^{T}(\mathbf{u},\mathbf{h}_{u}^{t-1})
\end{equation}
where $f_{e}^{T}$ is a temporal prediction model implemented as an LSTM \cite{Graves2012}, and $\mathbf{u}$ represents the learner. The expectation $\boldsymbol{\mu}_{u}^{t}$ and variance $\boldsymbol{\sigma}_{u}^{t}$ are then obtained via trainable weight matrices $\mathbf{W}$ and bias terms $\mathbf{b}$:
\begin{equation}
\boldsymbol{\mu}_{u}^{t}=\mathbf{W}_{\mu}\mathbf{h}_{u}^{t}+\mathbf{b}_{\mu}
\end{equation}
\begin{equation}
\boldsymbol{\sigma}_{u}^{t}=\text{Softplus}(\mathbf{W}_{\sigma}\mathbf{h}_{u}^{t}+\mathbf{b}_{\sigma})
\end{equation}
We posit that a high cognitive state should correlate with lower interaction uncertainty, and vice versa. We apply the variance to calibrate the expectation:
\begin{equation}
\mathbf{\hat{h}}_{u}^{t}=\boldsymbol{\mu}_{u}^{t}\odot(1-\boldsymbol{\sigma}_{u}^{t^{2}})
\end{equation}
This mechanism treats variance as a quantification of state reliability; when uncertainty is high, the mean is suppressed, whereas it is preserved when uncertainty is low. This results in a smoother representation that assists the diffusion model during de-noising.

\subsubsection{Goal-oriented Concept Encoder}
Existing methods often ignore the relationship between concepts and learning goals. Our goal-oriented concept encoder integrates target information to enhance personalization. Since structural relationships such as prerequisites significantly influence recommendations, we employ a self-attention mechanism to capture these associations.

We first project the initial concept embeddings $\mathbf{C}$ into query, key, and value spaces:
\begin{equation}
\mathbf{Q}=\mathbf{CW}_Q, \quad \mathbf{K}=\mathbf{CW}_K, \quad \mathbf{V}=\mathbf{CW}_V
\end{equation}
The scaled dot-product attention quantifies concept dependencies:
\begin{equation}
\hat{\mathbf{C}}=Softmax\left(\frac{\mathbf{QK}^{T}}{\sqrt{d}}\right)\mathbf{V}
\end{equation}
To preserve the original concept semantics, we apply a residual connection:
\begin{equation}
\hat{\mathbf{C}}_{a}=\mathbf{IV}+\hat{\mathbf{C}}
\end{equation}
Finally, the semantics of the learning goals $\mathbf{G}$ are aggregated and incorporated into the preliminary concept representations:
\begin{equation}
\mathbf{G}=\mathcal{\mathbf{W}}_{G}\left(\frac{1}{|\mathbf{G}|}\sum_{K=1}^{K}\mathbf{g}_{K}\right)+\mathbf{b}_{G}^{a}
\end{equation}
\begin{equation}
\hat{\mathbf{c}}_{b}=(\hat{\mathbf{C}}_{a}+\mathbf{G}+AvgPool(\hat{\mathbf{C}}_{a}))\mathbf{V}
\end{equation}
The term $AvgPool(\hat{\mathbf{C}}_{a})$ serves as a constraint to prevent the embeddings from deviating too far from the original concept space.

\subsection{Decoder}
The decoder's task is to predict the next concept representation based on the learner's cognitive state. Unlike discriminative methods that rely on pointer networks for selection, we utilize a generative diffusion model, which is better suited for generating idealized paths from scratch.

\subsubsection{State Update Auxiliary Component}
The reverse process of the diffusion model is iterative, spanning $T$ time steps to predict the latent representation of the next concept. To provide real-time guidance, we design an auxiliary component $f_{d}^{T}$ (implemented as a standard LSTM) to maintain and update the learner's cognitive state at each step:
\begin{equation}
\hat{\mathbf{h}}_{u}^{t}=f_{d}^{T}(\mathbf{u},\hat{\mathbf{h}}_{u}^{t-1})
\end{equation}
Since the generative diffusion scheme inherently accounts for historical uncertainty, a standard LSTM is sufficient for this component.

\subsubsection{Cognition-Adaptive Diffusion Model}
The reverse process starts from Gaussian noise and iteratively removes noise based on the real-time cognitive state $\mathbf{h}_{u}^{t}$. This is formalized as:
\begin{equation}
p_{\theta}(\tilde{\mathbf{c}}_{i_{t-1}}|\tilde{\mathbf{c}}_{i_{t}}) = \mathcal{N}(\tilde{\mathbf{c}}_{i_{t-1}}; \mu_{\theta}(\tilde{\mathbf{c}}_{i_{t}}, \mathbf{t}, f_{p}(\hat{\mathbf{h}}_{u}^{t})), \Sigma_{\theta}(\tilde{\mathbf{c}}_{i_{t}}, \mathbf{t}, f_{p}(\hat{\mathbf{h}}_{u}^{t})))
\end{equation}
The parameters are predicted by a noise network $\pi_{\theta}$. After $T$ iterations, we obtain the latent representation of the $i$-th concept. To map this back to a real concept, we calculate the matching degree $p_i$ using a transformation matrix $\mathbf{W}_d$ and apply a mask to avoid repeating visited concepts:
\begin{equation}
p_{i}=\mathbf{W}_{d}(\tilde{\mathbf{c}}_{i}+\mathbf{c})
\end{equation}
We use a softmax function to convert these degrees into a probability distribution for reinforcement learning sampling:
\begin{equation}
P_{i}(p_{i}|\hat{\mathbf{h}}_{u})=softmax(p_{i})
\end{equation}

\subsection{Training}
The optimization objective focuses on maximizing the goal mastery improvement rate $E_T$ from Equation (1). To ensure the predicted latent representations correspond to real concepts, we combine a reinforcement learning strategy gradient loss $\mathcal{L}_{r}$ with a noise prediction loss $\mathcal{L}_{d}$ as a penalty term, regulated by the weight $\lambda$:
\begin{equation}
\mathcal{L}=\mathcal{L}_{r}+\lambda\mathcal{L}_{d}
\end{equation}

The strategy gradient loss is defined as:
\begin{equation}
\mathcal{L}_{r}=-E_{T}\sum_{k=1}^{K}log P_{i}(p_{i}|\hat{\mathbf{h}}_{u})
\end{equation}
where $E_T$ is calculated using a Deep Knowledge Tracing (DKT) model.

The noise prediction loss $\mathcal{L}_{d}$ ensures the diffusion model accurately predicts the noise $\epsilon$ injected during the forward process, quantified by the mean squared error:
\begin{equation}
\mathcal{L}_{d}=||\epsilon-\pi_{\theta}(\tilde{\mathbf{c}}_{t}^{u},\mathbf{t},f_{p}(\hat{\mathbf{h}}_{u}^{t}))||^{2}
\end{equation}

\section{Experiments}

\subsection{Dataset Description}
To evaluate the effectiveness of the U-GLAD framework in learning path recommendation, we conducted experiments on three public educational datasets. The statistical information for these datasets is summarized in Table 1.

\begin{table}[htbp]
\centering
\caption{Statistics of the three educational datasets used in the experiments.}
\label{tab:dataset_stats}
\begin{tabular}{cccc}
\toprule
\textbf{Dataset} & \textbf{Students} & \textbf{Concepts} & \textbf{Avg. Seq. Length} \\ 
\midrule
Junyi         & 5,002 & 712  & 54.19 \\
SLP-Physics    & 663   & 1,451 & 54.24 \\
ASSISTments09 & 3,841 & 167  & 49.54 \\
\bottomrule
\end{tabular}
\end{table}

\begin{itemize}
    \item \textbf{Junyi}\cite{chang2015modeling}: Collected from Junyi Academy ( a Chinese e-learning platform), this dataset contains extensive records of learner-concept interactions and a rich knowledge graph reflecting structural concept relationships.
    \item \textbf{SLP-Physics}\cite{lu2021slp}: A benchmark dataset from the Smart Learning Partner (SLP) platform. We utilize the SLP-Physics subset, which covers three years of student interactions with physics concepts.
    \item \textbf{ASSISTments09}\cite{feng2009addressing}: Released by the ASSISTments online tutoring system, this dataset provides detailed information on student responses and exercise attempts. It is notably sparser compared to the other two datasets.
\end{itemize}

\subsection{Comparative Experiments}
We compared U-GLAD against three representative frameworks from the past three years to validate the effectiveness of our recommended paths:
\begin{itemize}
    \item \textbf{DLPR \cite{ZhangEtal2024}}: A difficulty-aware model that utilizes graph neural networks to aggregate prerequisite relations and employs reinforcement learning for path exploration.
    \item \textbf{SRC \cite{ChenEtal2023}}: Based on the Set-to-Sequence paradigm, this model uses self-attention to capture structural semantics and a pointer network for discriminative decoding.
    \item \textbf{LIGHT \cite{YuEtal2025}}: Enhances LPR by constructing prerequisite and synergistic graphs to enrich concept embeddings, also utilizing a pointer network for decoding.
\end{itemize}

\begin{table}[htbp]
\centering
\caption{Performance comparison (in terms of $E_{T}$) between U-GLAD and baseline models across three datasets with different path lengths.}
\label{tab:comparison_results}
\begin{tabular}{cccccc} 
\toprule
\textbf{Dataset} & \textbf{Path Length} & \textbf{SRC} & \textbf{DLPR} & \textbf{LIGHT} & \textbf{U-GLAD (Ours)} \\ 
\midrule
\multirow{3}{*}{ASSISTments09} & 10 & 0.4566 & 0.4788 & 0.4830 & \textbf{0.5117} \\
                               & 20 & 0.5416 & 0.5667 & 0.5656 & \textbf{0.6062} \\
                               & 30 & 0.5897 & 0.5854 & 0.5693 & \textbf{0.6125} \\
\addlinespace 
\multirow{3}{*}{SLP-Physics}    & 10 & 0.7826 & 0.7919 & 0.7143 & \textbf{0.8424} \\
                               & 20 & 0.8861 & 0.8957 & 0.8815 & \textbf{0.9210} \\
                               & 30 & 0.9101 & 0.9225 & 0.9126 & \textbf{0.9348} \\
\addlinespace
\multirow{3}{*}{Junyi}          & 10 & 0.4712 & 0.5562 & 0.5427 & \textbf{0.5871} \\
                               & 20 & 0.6554 & 0.6511 & 0.6633 & \textbf{0.6931} \\
                               & 30 & 0.6501 & 0.6656 & 0.6501 & \textbf{0.7024} \\
\bottomrule 
\end{tabular}
\end{table}

The recommendation performance across three path lengths (10, 20, 30) is presented in Table 2. U-GLAD consistently outperforms all baselines across all datasets and path length configurations.

The experimental results demonstrate that U-GLAD consistently and significantly outperforms all strong baseline models across all evaluation configurations. 
Notably, on the SLP-Physics dataset with a path length of 30, U-GLAD achieves a peak mastery rate ($E_T$) of $0.9348$, representing a substantial performance breakthrough compared to the latest state-of-the-art model, LIGHT ($0.9126$). 
Such comprehensive dominance validates the robust competitiveness of integrating uncertainty-aware encoding with generative diffusion decoding within complex educational environments. 

As the recommended path length scales from 10 to 30, a universal upward trend in $E_T$ is observed across all evaluated models, reflecting enhanced knowledge acquisition commensurate with increased learning engagement. 
Within this progression, U-GLAD maintains the steepest growth trajectory across different lengths. 
Particularly in short-path scenarios ($L=10$), U-GLAD exceeds LIGHT by approximately $12.8\%$ on the SLP-Physics dataset, suggesting its superior precision in identifying core concepts even under constrained path lengths. 

In contrast to baselines that rely on discriminative ranking via Pointer Networks, U-GLAD directly predicts the latent representation of the next concept through a cognition-adaptive diffusion model. 
This generative approach allows for the reconstruction of an idealized learning trajectory in the latent space, ultimately yielding learning paths of higher pedagogical quality.

\subsection{Ablation Study}
To verify the indispensable role of each component within U-GLAD, we designed four variants for ablation analysis:
\begin{itemize}
    \item \textbf{U-GLAD w/o Uncertainty-aware}: Replaces the Gaussian LSTM with a standard LSTM to evaluate the impact of mitigating historical interaction uncertainty on cognitive state modeling.
    \item \textbf{U-GLAD w/o Diffusion}: Replaces the generative diffusion model with a discriminative MLP mapping, testing the necessity of generative decoding for predicting next-concept latent representations.
    \item \textbf{U-GLAD w/o Noise Prediction Loss}: Trains the model without the noise prediction loss penalty, exploring whether the diffusion model generates "hallucinated" or invalid latent representations.
    \item \textbf{U-GLAD w/o Goal-oriented}: Removes the objective-aware encoding capability to test the necessity of tailoring concept representations to specific learning goals.
\end{itemize}

As shown in Figure 2, all variants exhibited performance degradation. We conclude that: (1) historical uncertainty negatively impacts path quality, which our Gaussian LSTM effectively mitigates; (2) generative decoding produces more accurate latent representations than discriminative methods; (3) the noise prediction loss is a critical constraint for generating valid concepts; and (4) goal-aware personalization is vital for effective recommendations.

\begin{figure}[H]
\centering
\includegraphics[width=0.90\textwidth]{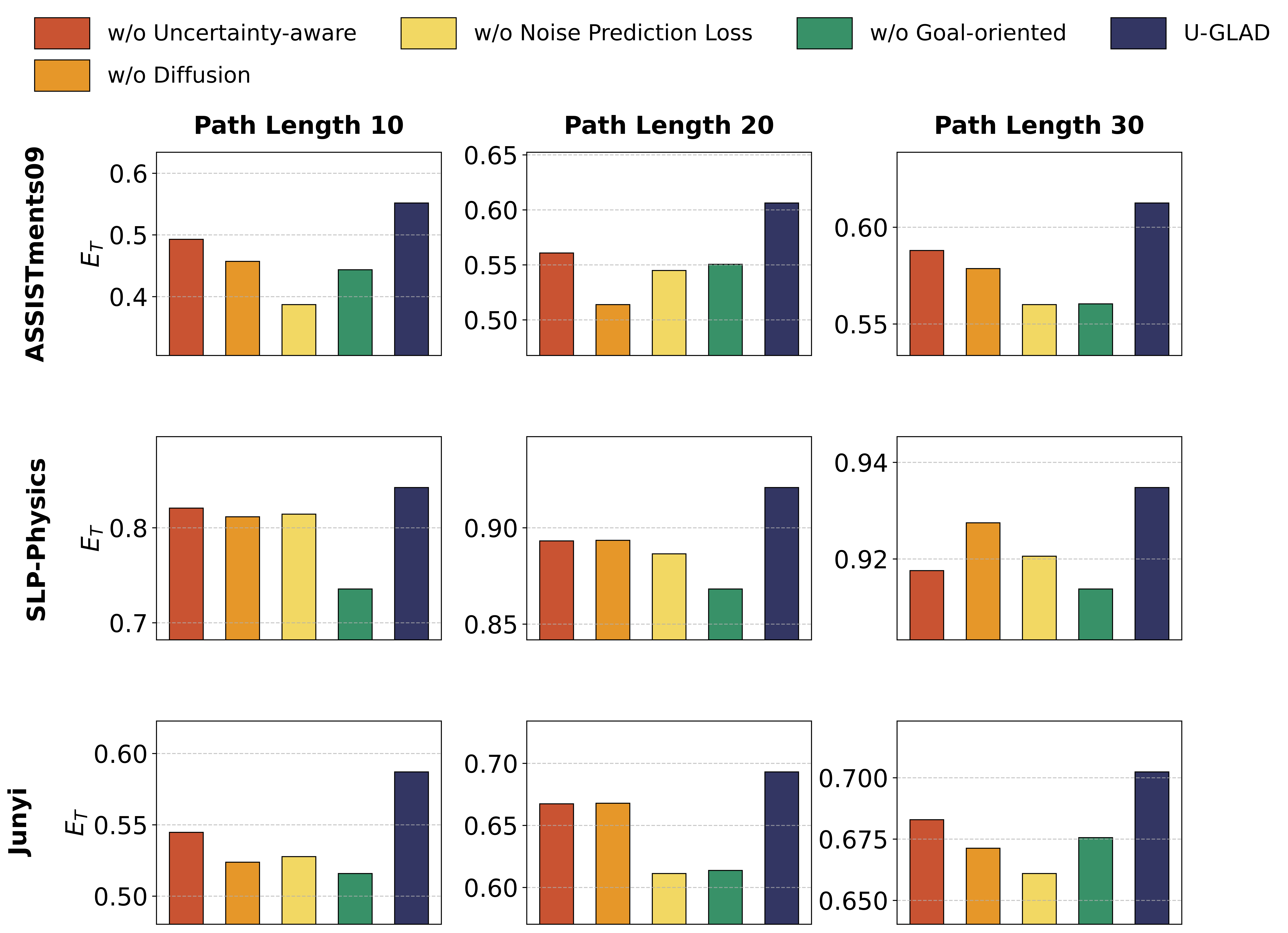} 
\caption{Visualization of ablation studies across three datasets. We evaluate the contribution of four core components: (1) w/o Uncertainty-aware (replacing Gaussian LSTM with standard LSTM), (2) w/o Diffusion (generative vs. discriminative decoding), (3) w/o Noise Prediction Loss (absence of noise prediction penalty), and (4) w/o Goal-oriented (removing objective-aware encoding).}
\label{fig:ablation}
\end{figure}

\subsection{Cognitive State De-noising Analysis}
The Gaussian LSTM is designed to form robust cognitive state representations by attenuating interaction noise. To quantify this, we adopt the \textbf{State Instability} metric ($s$), which measures the Euclidean distance between cognitive state vectors at adjacent time steps:
\begin{equation}
s=\frac{1}{T-1}\sum_{t=1}^{T-1}||\mathbf{h}_{u}^{t+1}-\mathbf{h}_{u}^{t}||_{2}
\end{equation}

\begin{figure}[H]
\centering
\includegraphics[width=0.90\textwidth]{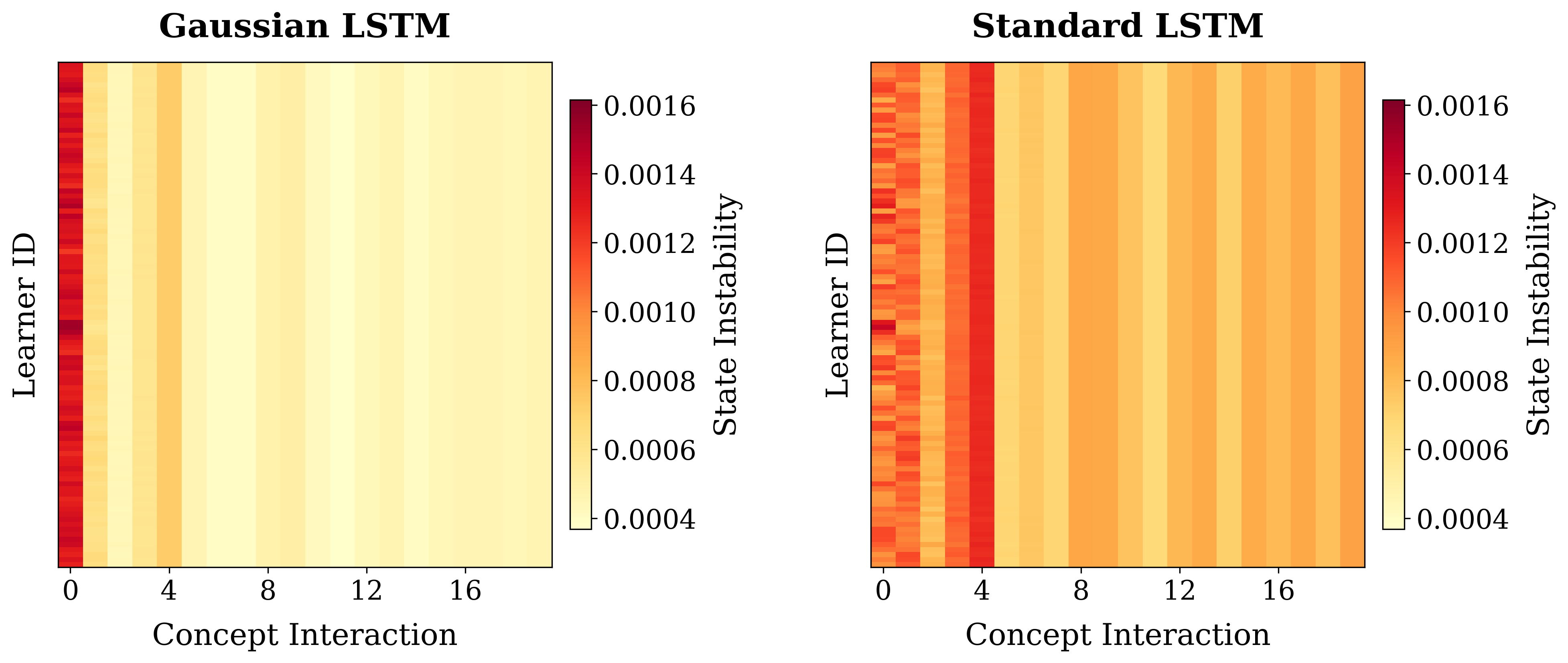} 
\caption{Cognitive state instability analysis. The heatmaps visualize the state instability metric ($s$) for Gaussian LSTM (left) and the standard LSTM (right) on the Junyi dataset. The results demonstrate that while both models exhibit instability initially, Gaussian LSTM effectively learns to suppress historical interaction uncertainty as more data is ingested, resulting in smoother and more stable cognitive state representations compared to the persistent fluctuations of the standard LSTM.}
\label{fig:smoothness}
\end{figure}

Visualized in the heatmaps of Figure 3, the Gaussian LSTM (left) initially exhibits higher instability but quickly learns to suppress uncertainty as more data is ingested, resulting in a smoother and more stable representation compared to the standard LSTM (right), which shows persistent fluctuations.

\subsection{Hyperparameter Sensitivity Analysis}
We explored the impact of the diffusion model's reverse iterations ($T$) and the noise loss weight ($\lambda$) with a fixed path length $L=20$.

\begin{figure}[H]
    \centering
    \includegraphics[width=\textwidth]{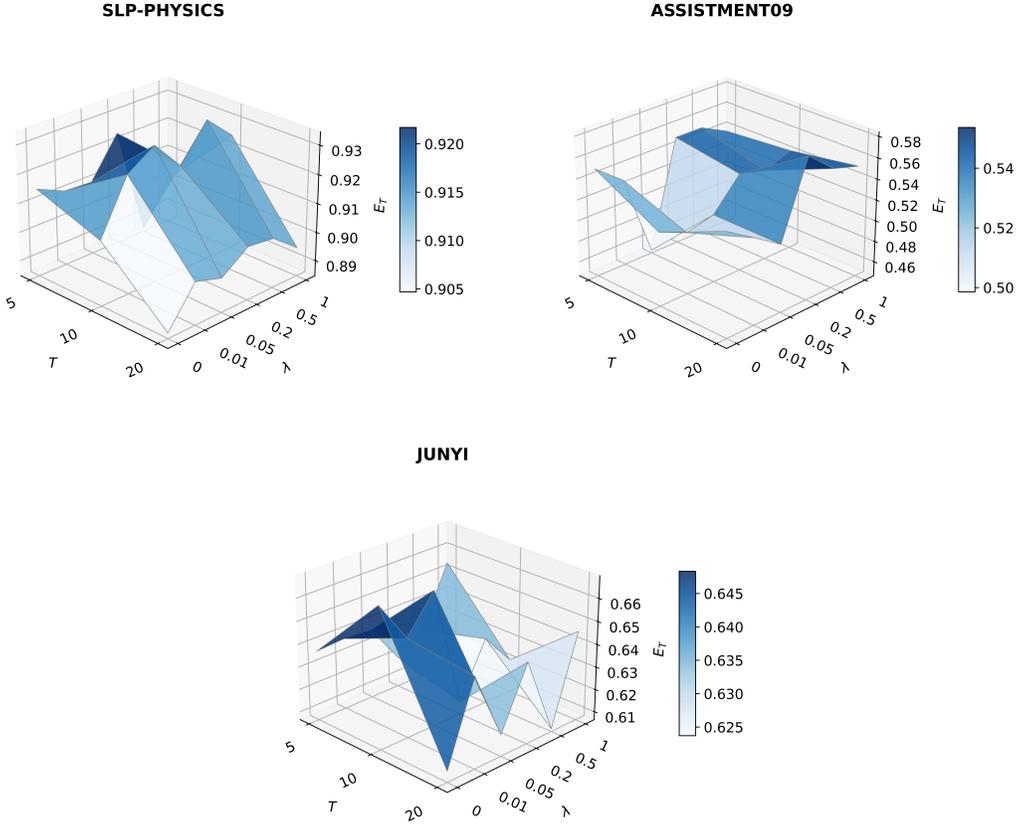}
    \caption{Impact of the number of diffusion model reverse process iterations $T$ and noise prediction loss weight $\lambda$ on the U-GLAD framework.}
    \label{fig:hyperparameter_sensitivity}
\end{figure}

On the SLP-Physics and Junyi datasets, performance typically peaks at $T=10$. Excessive iterations lead to over-fitting of historical sequences, which may not represent the ideal path for maximizing $E_T$. On the sparse ASSISTments09 dataset, a U-shaped trend was observed. At $T=5$, the decoder likely performs a beneficial "vague matching" optimized by strategy gradients. At $T=10$, insufficient learning of the noise distribution leads to latent representations that deviate from real concepts. Performance recovers at $T=20$ as the noise network better characterizes the underlying data distribution.

\section{Conclusion}
This work investigates how to mitigate the negative impact of historical interaction uncertainty on cognitive state modeling and introduces a generative diffusion approach to learning path recommendation. The proposed U-GLAD framework demonstrates consistent superiority across multiple benchmarks. Our Gaussian LSTM provides robust, de-noised cognitive state representations, while the goal-oriented encoder enhances personalization. As one of the early attempts to integrate generative diffusion decoding into LPR, this study validates the potential of diffusion models for educational sequence planning.

\section*{CRediT authorship contribution statement}

\textbf{Xiangrui Xiong:} Writing – original draft, Methodology, Software, Validation. \textbf{Hang Liang:} Writing – review \& editing. \textbf{Zifei Pan:} Data curation. \textbf{Baiyang Chen:} Supervision, Writing – review \& editing, Visualization. \textbf{Yanli Lee:} Supervision, Writing – review \& editing, Funding acquisition.

\section*{Declaration of Competing Interest}

The authors declare that they have no known competing financial interests or personal relationships that could have appeared to influence the work reported in this paper.

\section*{Data availability}
Data will be made available on request.





\bibliographystyle{elsarticle-num}
\bibliography{cas-refs}
\end{document}